\documentclass[aps,prd,secnumarabic,amssymb, amsmath,nobibnotes,nofootinbib,11pt]{revtex4}
\usepackage{amsfonts,amsmath,hyperref,url, color}
\usepackage{eurosym}

\def\bbbone{{\mathchoice {\rm 1\mskip-4mu l} {\rm 1\mskip-4mu l}
{\rm 1\mskip-4.5mu l} {\rm 1\mskip-5mu l}}}

\def\he{\hat e}

\begin{document}

\title{%
A short introduction to $\kappa$-deformation}
\author{J. Kowalski-Glikman}
\email{jerzy.kowalski-glikman@ift.uni.wroc.pl}\affiliation{Institute for
Theoretical Physics, University of Wroc\l{}aw, Pl.\ Maksa Borna 9,
Pl--50-204 Wroc\l{}aw, Poland}
\date{\today}

\begin{abstract}
In this short review we describe some aspects of $\kappa$-deformation. After discussing the algebraic and geometric approaches to $\kappa$-Poincar\'e algebra we construct the free scalar field theory, both on non-commutative $\kappa$-Minkowski space and on curved momentum space. Finally, we make a few remarks concerning interacting scalar field.
\end{abstract}
\maketitle

\section{Introduction}

Poincar\'e algebra is one of the most important structures of modern high energy physics. On the one hand it is an algebra of infinitesimal symmetries of flat Minkowski spacetime, including rotations with generators $M_i$, boosts with generators $N_i$, and translations with generators $P_0$, $P_i$, $i=1,2,3$ satisfying the following commutator algebra $\kappa$-Poincar\'e is a quantum deformation of Poincar\'e algebra\footnote{We use the $(-,+,\ldots,+)$ signature.}
\begin{equation}\label{i1}
  [M_i, P_j] = i\, \epsilon_{ijk} P_k, \quad [M_i, P_0] =0
\end{equation}
\begin{equation}\label{i2}
   \left[N_{i}, {P}_{j}\right] = i\,  \delta_{ij} \, P_0
 \, ,
\quad
  \left[N_{i},P_0\right] = i\, P_{i}.
\end{equation}
\begin{equation}\label{i3}
   [M_i, M_j] = i\, \epsilon_{ijk} M^k, \quad    [M_i, N_j] = i\, \epsilon_{ijk} N^k, \quad    [N_i, N_j] =- i\, \epsilon_{ijk} N^k, \quad
\end{equation}
As it is well known elementary particles can be thought of as representations of this algebra and this is the reason why Poincar\'e algebra plays such a fundamental role in modern physics. In the language of quantum field theory we demand that the action is Poincar\'e invariant, which is achieved by using the fields that transform covariantly under relevant representations of the Poincar\'e group. As for the states, they must transform under unitary representation of this group. It usually comes without saying that when acting on the states that have the form of tensor products (for examply, the multiparticle states in Fock representation), the action of Poincar\'e group is Leibnizean, i.e., the generator acts on the first state leaving the other intact, then on the second, etc, and then the results of these actions are summed. As we will see this is Leibniz rule that can be nontrivially generalized and a particular generalization is the essence of $\kappa$-deformation.

It was about 25 years ago when such generalization was proposed in a series of seminal papers \cite{Lukierski:1991pn}, \cite{Lukierski:1992dt}, \cite{Lukierski:1993wxa}, in which a Hopf algebra, being a deformation of Poincar\'e algebra was derived by contraction the Hopf algebra $SO_q(3,2)$ (deformed Anti de Sitter algebra.) It turned out that the deformation parameter of the resulting algebra, denoted by $\kappa$, is by construction dimensionful, with dimension of mass. Since Poincar\'e symmetry is a symmetry of flat spacetime it was claimed from the very beginning that its deformation, $\kappa$-Poincare symmetry, must have something to do with quantum gravity, and therefore it is expected that the parameter $\kappa$ should be identified with the quantum gravity energy scale, the Planck mass, $M_{Pl}\sim 10^{19}$ GeV. It was only recently when this claim found its solid proof (albeit still in the framework of a rather restricted model -- see below.)

Since the deformed $\kappa$-Poincar\'e algebra possesses a mass scale, a physical theory, in which this algebra plays a role similar to that the standard Poincar\'e symmetry plays in the theory of particles and fields, will also have to incorporate this scale in some way. One of the ways one can incorporate the scale of mass in relativistic theory is to assume that it plays a role of a second observer independent scale, in addition to velocity of light, but built in the theory in such a way that the relativity principle (equivalence of internal observers) still holds. This idea lead to the formulation of Doubly Special Relativity \cite{Amelino-Camelia:2000ge}, \cite{Amelino-Camelia:2000mn}, \cite{KowalskiGlikman:2001gp}, \cite{Bruno:2001mw} (see \cite{Kowalski-Glikman:2004qa} for review.)

The Poincar\'e algebra can be seen as describing the flat momentum space and its Lorentz symmetry. Having it, one can construct the dual object, the flat Minkowski space with the Lorentz group action on it. In the deformed case the situation is analogous, but much more interesting. First of all, the space dual to the momentum sector of $\kappa$-Poincar\'e is a non-commutative space, with the non-commutativity scale $\ell$ equal to the inverse of $\kappa$ and thus of order of the Planck length, $\ell \sim 10^{-35}$ m \cite{kappaM1}, \cite{kappaM2}. Second, the geometry of momentum space becomes non-trivial: the momentum space associated with $\kappa$-Poincar\'e has a form of the manifold of the group $AN(3)$, being a submanifold of de Sitter space with constant curvature proportional to $\kappa^2$ \cite{KowalskiGlikman:2002ft}, \cite{KowalskiGlikman:2003we}, \cite{KowalskiGlikman:2004tz}. As a consequence of the non-trivial geometry  of momentum space one is forced to abandon one of the basic postulates of modern physics, the absolute locality, and to replace it with the ``principle of relative locality'' \cite{AmelinoCamelia:2011bm}, \cite{AmelinoCamelia:2011pe}, \cite{AmelinoCamelia:2011nt}.

As said above, it was very clear from the early days of $\kappa$-deformation that it has to have something to do with quantum gravity. Surprisingly, the direct link was hard to establish, despite numerous attempts \cite{AmelinoCamelia:2003xp}, \cite{Freidel:2003sp}, \cite{KowalskiGlikman:2008fj}. Only recently the direct proof of emergence of $\kappa$-deformation from quantum gravity appeared \cite{Cianfrani:2016ogm}. In this paper, in the context of $2+1$ Euclidean quantum gravity it was shown that $\kappa$-Poincar\'e is a symmetry of flat, quantum spacetime. Let us now briefly recall how this result came about.

The starting point of the paper  \cite{Cianfrani:2016ogm} is to define the meaning of the term ``symmetry of flat quantum spacetime.'' We know that in classical general relativity the flat spacetime is Minkowski space and its symmetries are given by (\ref{i1}-\ref{i3}). These Poincar\'e algebra commutational relations can be derived from classical gravity as follows. In the Hamiltonian treatment of general relativity, the theory is completely described by two infinite, labeled by space point $x$ sets of constraints, the diffeomorphism constraints ${\cal D}_i(x)$, generating diffeomorphisms in space and  the Hamiltonian constraint ${\cal H}(x)$ related to diffeomorphism in time direction. Instead of working with constraints labeled by points one can use their smeared version, where the constraints are integrated against arbitrary functions, to wit
$$
{\cal D}[\vec{f}]=\int_\Sigma{\cal D}_i(x) f^i(x)\,,\quad {\cal H}[g]=\int_\Sigma{\cal H}(x) g(x)\,.
$$
Then the smeared constraints satisfy the Poisson bracket algebra
\begin{align}
&\big\{{\cal D}[f_1],\, {\cal D}[f_2]\big\}={\cal D}\big[[f_1,\, f_2]\big]\nonumber \\
&\big\{{\cal D}[f],\, {\cal H}[g]\big\}={\cal H}[f^i\partial_ig]\nonumber\\
&\big\{{\cal H}[g_1],\, {\cal H}[g_2]\big\}={\cal D}[f(g_1,g_2)]\label{i4}
\end{align}
with
\begin{equation}\label{i5}
[f_1,\, f_2]=f_1^i\,\partial_i\vec{f}_2-f_2^i\,\partial_i\vec{f}_1\qquad
f^i(g_1,g_2)=h^{ij}(g_1\,\partial_jg_2-g_2\,\partial_jg_1)\,
\end{equation}
where $h^{ij}$ is the metric on the three-dimensional space $\Sigma$. Remarkably, the algebra (\ref{i4}) reproduces the Poincar\'e algebra if the smearing functions are chosen such that the vector $\xi^\mu=(h, f^i)$ is one of the Killing vectors of Minkowski space and $h^{ij}$ is the flat Euclidean metric\footnote{The analogous statement holds for other maximally symmetric space, de Sitter or Anti de Sitter space.}. Explicitly, we represent translation as
$$
P_0 = \int_\Sigma {\cal H}\,,\quad P_i =\int_\Sigma {\cal D}_i\,,
$$
while for the Lorentz generators we have
$$
N_i = \int_\Sigma x_i {\cal H}\,,\quad M_i =\epsilon_{ijk}\, \int_\Sigma x^j\, {\cal D}^k\,,
$$
and one can check that after substituting this to (\ref{i4}) one obtains the Poincar\'e algebra.
The idea now is to find the commutator algebra of quantum operators, corresponding to the classical diffeomorphism and Hamiltonian constraints, use again the smearing functions being the Killing vectors of Minkowski space and the Euclidean metric in the appropriate places, and finally, investigate the properties of the resulting algebra of symmetries of quantum flat space.

Unfortunately, for technical reasons this program cannot be carried on in the physical 3+1 dimensions. However, one can do that in the case of Euclidean gravity with positive cosmological constant in 2+1 dimensions, where the algebra of symmetries of ``quantum Euclidean de Sitter space'' turns out to be a direct sum of two Hopf algebras $su_q(2)\oplus su_{-q}(2) \sim so_q(4)$, where the deformation parameter $q=\exp(i\hbar\sqrt\Lambda/2\kappa)$, with $\Lambda$ being the cosmological constant, and $\kappa$ the three-dimensional Newton's constant (of dimension of mass.) One can then take the limit of vanishing cosmological constant (which turns out to be a quite non-trivial procedure) obtaining as a result the algebra of symmetries of ``quantum flat spacetime'' which turns out to be the (three dimensional) $\kappa$-Poincar\'e algebra. The reader is referred to the paper  \cite{Cianfrani:2016ogm} where all the details of this construction are described.

The fact that the symmetries of ``flat quantum spacetime'' in three dimensions are $\kappa$-deformed indicates that the same should also happen in the physical 3+1 dimensions. The brief argument in favor of this claim goes as follows (for more elaborate version see \cite{Freidel:2003sp}.) Classically 2+1 dimensional Minkowski space is a submanifold of the 3+1 dimensional one, and therefore the 2+1 dimensional Poincar\'e algebra is a subalgebra of the 3+1 dimensional one. Also the algebra of constraints of 2+1 gravity is a subalgebra of the 3+1 dimensional one (one can use only the class of smearing functions satisfying $f^i =(f^1, f^2, 0)$, with $f^1, f^2, g$ depending only on two coordinates on $\Sigma$, $(x^1, x^2)$.) Then it seems natural to expect that the algebra of commutators of quantum constraint operators of 3+1 quantum gravity contains as a subalgebra the 2+1 dimensional one. But the latter is a $\kappa$-deformed algebra and therefore the algebra of symmetries of 3+1 dimensional quantum flat space cannot be just the ordinary Poincar\'e algebra. Even more, since as we will see the rotation sector of $\kappa$-deformed algebra is un-deformed, the 2D vector $f^i =(f^1, f^2, 0)$ can be rotated to the 3D one in the un-deformed way, and therefore the resulting deformed 3+1 dimensional symmetry algebra should be also the $\kappa$-Poincar\'e algebra in 3+1 dimensions.

The present review of $\kappa$-deformation is complementary to the one I wrote five years ago \cite{Kowalski-Glikman:2013rxa}, and I concentrate here on formal structures of $\kappa$-Poincar\'e algebra and the dual $\kappa$-Minkowski space. In the next section the Hopf algebra perspective on $\kappa$-deformation is described. Then in the following section the complementary, more group theoretical and geometrical picture is presented. The technical tools described earlier are then used to formulate the free scalar field theory. Finally, the interacting scalar field and path integral are briefly described.

\section{Hopf algebra perspective}

There are several equivalent ways one can introduce $\kappa$-deformation. Let us start with the most direct one, for which the starting point is the $\kappa$-Poincar\'e algebra, as derived by Lukierski, Nowicki, Ruegg, and Tolstoy \cite{Lukierski:1991pn}, \cite{Lukierski:1992dt}, \cite{Lukierski:1993wxa} and presented in its final form by Majid and Ruegg \cite{kappaM2}. This is this final form that we will present below

Being a Hopf algebra, the $\kappa$-Poincar\'e algebra consists of several mutually compatible structures, the algebra, the co-algebra, and the antipode. Their physical relevance of these structures will become clear from what will be said below.
\newline

{\em The algebra}. The algebra of $\kappa$-Poincar\'e can be thought of as an universal enveloping (nonlinear) algebra with the generators of boosts $N_i$, rotations $M_i$, energy $k_0$, and linear momentum $k_i$ satisfying the following commutational relations
\begin{equation}\label{1}
  [M_i, k_j] = i\, \epsilon_{ijk} k_k, \quad [M_i, k_0] =0
\end{equation}
\begin{equation}\label{2}
   \left[N_{i}, {k}_{j}\right] = i\,  \delta_{ij}
 \left( {\kappa\over 2} \left(
 1 -e^{-2{k_0}/\kappa}
\right) + {{\mathbf{k}^2}\over 2\kappa}  \right) - i\,\frac1{\kappa}\, k_{i}k_{j} ,
\quad
  \left[N_{i},k_0\right] = i\, k_{i}.
\end{equation}
\begin{equation}\label{3}
   [M_i, M_j] = i\, \epsilon_{ijk} M^k, \quad    [M_i, N_j] = i\, \epsilon_{ijk} N^k, \quad    [N_i, N_j] =- i\, \epsilon_{ijk} N^k, \quad
\end{equation}
The basis of $\kappa$-Poincar\'e algebra defined above is called the `bicrossproduct' basis.

The most important thing to notice about this algebra is that it is a {\em deformation} of the standard Poincar\'e algebra, which goes away in the limit $\kappa\rightarrow\infty$, i.e., in this limit the algebra (\ref{1}--\ref{3}) becomes the standard Poincar\'e algebra (\ref{i1})--(\ref{i3}). The second important point is that that the deformation parameter $\kappa$ has the dimension of mass and therefore, since the boost and rotation generators are dimensionless, only the momentum sector is being deformed.

One can ask a question if it is possible to simplify this algebra by a non-linear transformation of the generators, and the answer is affirmative. In fact in terms of the new momenta generators $P_\mu$
\begin{eqnarray}
 {P_0}(k_0, \mathbf{k}) &=&  \kappa\sinh
{\frac{k_0}{\kappa}} + \frac{\mathbf{k}^2}{2\kappa}\,
e^{  \frac{k_0}{\kappa}} \nonumber\\
 P_i(k_0, \mathbf{k}) &=&   k_i \, e^{  \frac{k_0}{\kappa}}
 \label{4}
\end{eqnarray}
the algebra (\ref{1}--\ref{3}) becomes the standard Poincar\'e algebra. This basis of the algebra is called `classical' and is discussed in details in \cite{Borowiec:2009vb}, where its subtle mathematical aspects are emphasized.

In what follows we will also need the object
\begin{equation}\label{5}
    {P_4}(k_0, \mathbf{k}) =  \kappa\cosh
{\frac{k_0}{\kappa}} - \frac{\mathbf{k}^2}{2\kappa}\,
e^{  \frac{k_0}{\kappa}}
\end{equation}
which will find its natural place in the physical applications of $\kappa$ deformation.

The set of momentum generators like $k_\mu$ and $P_\mu$ defines what is called in the literature {\em the basis} of the algebra. Clearly there are infinitely many different bases; the only requirement is that when the deformation is being switched off, $\kappa\rightarrow\infty$, all of them turn back to the standard Poincar\'e algebra.

From the physical perspective therefore it should not matter, which basis one uses to describe physics. In fact there is no physical principle that tells us that one particular basis is preferred by nature -- they should play a role similar to that played by coordinate systems in general relativity, where relativity principle tells us that descriptions of physical phenomena in any coordinates are equivalent. As we will see later this analogy is in fact more direct than it seems to be.

On the other hand, different bases look really differently (compare for example (\ref{1}--\ref{3}) with the standard Poincar\'e algebra), so the question arises how is it possible that they describe the same physics. Moreover, since among available bases there is the classical one with the algebra identical to the standard Poincar\'e one, is there some new physics described by the deformed theory? There must be some more structures involved that make the $\kappa$ deformation differ from the standard un-deformed physics.
\newline

{\em The coproduct}. One of the elements of this larger structure is the coproduct, defining the way the generators of the algebra act on products; for example, the co-product of the energy $k_0$ defines the way in which one can compute the total energy of two- and many-particles states. In the abstract terms the coproduct $\Delta$ of is a mapping from the algebra ${\cal A}$ in question to the tensor product of two copies of the algebra ${\cal A}\otimes {\cal A}$.

In $\kappa$-deformation the co-product of rotation generators is un-deformed, meaning that it acts in the standard, Leibnizean way on tensor products
\begin{equation}\label{6}
  \Delta( M_i) = M_i\otimes \bbbone + \bbbone \otimes M_i
\end{equation}
For boosts the co-product is more complicated and reads
\begin{equation}\label{7}
    \Delta (N_i)=N_i\otimes \bbbone +e^{-{k_0}/\kappa}\otimes N_i+\frac1\kappa\, \epsilon_{ijk}k_j\otimes M_k.
\end{equation}
while for momentum generators we have
 \begin{equation}\label{8}
   \Delta(k_i) = k_i \otimes \bbbone + e^{-k_0/\kappa} \otimes k_i, \quad \Delta(k_0) = k_0 \otimes \bbbone + \bbbone \otimes k_0
\end{equation}

To see what the co-product is needed for consider a two-particle state, in which the first particle carries the (four) momentum $k^{(1)}$, while the second $k^{(2)}$; then, the corresponding two particle state is given by
 $$
 \left|k\right> = \left|k^{(1)}\right>\otimes \left|k^{(2)}\right>\,,\quad k^{(total)}_\mu \left|k\right> = \Delta(k)_\mu\, \left|k^{(1)}\right>\otimes \left|k^{(2)}\right>
  $$
so that the total momentum of this state is
$$
k^{(total)}_0 = k^{(1)}_0+k^{(2)}_0\,,\quad k^{(total)}_i = k^{(1)}_i+e^{-k^{(1)}_0/\kappa}\, k^{(2)}_i
$$

Among many things that the coproduct is good for a very interesting one is the construction of the non-commutative spacetime associated with the $\kappa$-Poincar\'e algebra \cite{kappaM2}, . It goes as follows. Similarly to the standard Poincar\'e case we define the spacetime as a dual to the momentum sector of the algebra (\ref{1})--(\ref{3}). In the first step we introduce the position variables $X^\mu$ and the action of momenta and Lorentz generators  on them
\begin{equation}\label{9}
k_\nu\vartriangleright X^\mu=-i \delta^\mu_\nu\,,\quad N_i\vartriangleright X^j =i X^0\, \delta_{i}^j\,, \quad N_i\vartriangleright X^0 =i X^i
\end{equation}
Next we have to define the action of momenta on polynomials built from $X$. This is the step, in which the coproduct finds its natural application, to wit
\begin{equation}\label{10}
  k\vartriangleright (X^\mu X^\nu) \equiv \sum k_{(1)} \vartriangleright (X^\mu)  \, k_{(2)}\vartriangleright (X^\nu)
\end{equation}
where we made use of the, so called, Sweedler notation
$$
\Delta k = \sum k_{(1)}\otimes k_{(2)}
$$
Then, from (\ref{10}) we compute
$$
k_0\vartriangleright (X^0 X^i) = k_{0} \vartriangleright (X^0)  \, \bbbone \vartriangleright (X^i ) +\bbbone \vartriangleright (X^0)  \, k_{0} \vartriangleright (X^i ) = -i X^i
$$
and
$$
k_0\vartriangleright (X^i X^0) = k_{0} \vartriangleright (X^i)  \, \bbbone \vartriangleright (X^0 ) +\bbbone \vartriangleright (X^i)  \, k_{0} \vartriangleright (X^0 )=-i X^i
$$
Similarly
$$
k_j\vartriangleright (X^0 X^i) = k_{j} \vartriangleright (X^0)  \, \bbbone \vartriangleright (X^i ) +e^{-k_0/\kappa} \vartriangleright (X^0)  \, k_{j} \vartriangleright (X^i ) = -i\delta_j^i +\frac1\kappa\, \delta_j^i
$$
and
$$
k_j\vartriangleright (X^i X^0) = k_{j} \vartriangleright (X^i)  \, \bbbone \vartriangleright (X^0 ) +e^{-k_0/\kappa} \vartriangleright (X^i)  \, k_{j} \vartriangleright (X^0 )=-i \delta_j^i
$$
We see therefore that
$$
k_0\vartriangleright [X^0, X^i] =0\,,\quad k_j\vartriangleright [X^0, X^i] =\frac1\kappa\, \delta_j^i
$$
and therefore we conclude that the positions do not commute
\begin{equation}\label{11}
  \left[X^0, X^i\right]= \frac{i}{\kappa}\, X^i
\end{equation}
One can check that all other commutators of positions vanish.

The non-commutative spacetime defined by relations (\ref{11}) is called $\kappa$-Minkowski space.

It is worth noticing that since the momenta commute, the same calculation done for polynomials of momenta forces us to assume that the coproduct of positions is un-deformed
\begin{equation}\label{12}
  \Delta(X^\mu) = X^\mu \otimes \bbbone + \bbbone \otimes X^\mu\,.
\end{equation}

The commutational relation (\ref{11}) does not look covariant, but in fact it is, as a result of a nontrivial boosts' coproduct. Let us compute using (\ref{7})
$$
N_j\vartriangleright [X^0, X^i] = -\frac1\kappa\, X^0\, \delta_j^i = N_j\vartriangleright \frac{i}{\kappa}\, X^i\,,
$$
where $N_j\vartriangleright X^0=i X_j$, $N_j\vartriangleright X^j=i\delta_j^i X_0$ which indeed shows that  $\kappa$-Minkowski space defining commutator is Lorentz-covariant.

When one changes the basis of the algebra, the coproduct changes too, in a natural way. For example, in the case of the classical basis (\ref{4}) we have
\begin{align}
\Delta P_0 &=  \kappa\sinh
{\frac{\Delta k_0}{\kappa}} + \frac{\Delta\mathbf{k}^2}{2\kappa}\,
e^{  \frac{\Delta k_0}{\kappa}}\nonumber\\
&= \frac1\kappa\, P_{0}\otimes (P_{0}+P_{4})+\sum P_{k}(P_{0}+P_{4})^{-1}\otimes P_{k}
 +\kappa\,(P_{0}+P_{4})^{-1}\otimes P_{0} \nonumber\\
\Delta  P_i&=  \Delta k_i \, e^{  \frac{\Delta k_0}{\kappa}} =\frac1\kappa\, P_{i}\otimes (P_{0}+P_{4}) +\bbbone\otimes P_{i}
 \label{13}
 \end{align}
In deriving these relations one uses the identities $\Delta \mathbf{k}^2 = \Delta \mathbf{k}\, \Delta \mathbf{k}$,  $e^{  {\Delta k_0}/{\kappa}}= e^{k_0/\kappa}\otimes e^{k_0/\kappa}$, and $e^{k_0/\kappa} = P_0+P_4$.

It is important to notice that the coproduct of $\kappa$-Poincar\'e algebra is not symmetric, and for this reason it is non-reducibly not trivial in any basis. This property distinguishes the $\kappa$-Poincar\'e algebra from the naive deformations of the Poincar\'e algebra.

To see this more explicitly, let us consider the standard Poincar\'e algebra
\begin{align}
  [M_i, p_j] &= i\, \epsilon_{ijk} p_k\,, \quad [M_i, p_0] =0 \nonumber\\
   \left[N_{i}, {p}_{j}\right] &= i\,  \delta_{ij}\, p_0
\,,
\quad
  \left[N_{i},p_0\right] = i\, p_{i}\nonumber\\
   [M_i, M_j] &= i\, \epsilon_{ijk} M^k, \quad    [M_i, N_j] = i\, \epsilon_{ijk} N^k, \quad    [N_i, N_j] =- i\, \epsilon_{ijk} N^k,\nonumber
\end{align}
with coproducts of all generators trivial, $\Delta(p_0) = p_0 \otimes \bbbone + \bbbone \otimes p_0$, etc. One can convince oneself that if we apply the transformation (\ref{4}) to find out what is the coproduct of momentum variables $k_\mu$ the resulting coproduct will be quite complicated, but still symmetric. Then, proceeding as we did above, one can check that this new coproduct is compatible with commutative spacetime structure. In fact any symmetric coproduct can be turned to the trivial one by an appropriate change of the algebra generators.

One can therefore conclude that in the case of $\kappa$-Poincar\'e {\em it is not the deformation of the algebra that really matters, but the coproduct and the associated non-commutative spacetime structure.}
\newline

There is yet another interesting calculation which requires the knowledge of the coproduct. In order to define the whole of the phase space of a $\kappa$-deformed system, we need the commutators generalizing the Heisenberg algebra. These are defined with the help of the so-called Heisenberg double \cite{AmelinoCamelia:1997jx}.
\begin{equation}\label{14}
  [k, X] =\sum X_{(1)} \left<k_{(1)} , X_{(2)} \right> k_{(2)}-Xk
\end{equation}
where we made use of the pairing between momenta and positions
\begin{equation}\label{15}
  \left<k_\mu , X^\nu \right> = -i\delta_\mu^\nu\,,\quad \left<\bbbone , \bbbone \right> =1
\end{equation}
so for example
\begin{equation}\label{16}
  [k_0, X^0] = -i\,, \quad  [k_i, X^j] = -i\,\delta_i^j \,, \quad [k_i,X^0]= \frac i\kappa k_i\,.
\end{equation}
One can check that these deformed Heisenberg relations together with (\ref{11}) satisfy Jacobi identities and are covariant with respect to Lorentz transformations. It should be remarked at this point that the phase space as a whole does not have the Hopf algebra structure. In order to deform the phase space, one presumably has to make use of more general structures, like the one of Hopf algebroid (see \cite{Lukierski:2015zqa} for further discussion).

This concludes our discussion of the coproduct and now let us turn to the last structure on the list.
\newline

{\em Antipode}, denoted by $S$, is a deformed minus and is given by
\begin{equation}\label{17}
  S(M_i) = - M_i\,,\quad S(N_i) =  -e^{{k_0}/\kappa}\left(N_i-\frac1\kappa\, \epsilon_{ijk}k_jM_k\right)
\end{equation}
while for momenta we have
\begin{equation}\label{18}
  S(k_{0})=-k_{0}\,,\quad S(k_{i})=-k_{i}e^{k_{0}/\kappa}
\end{equation}
The physical meaning of the antipode will become clear from the more group theoretical and geometrical perspective that we are going to discuss now.

\section{Group theoretical/geometrical perspective}

Let us now try to look at $\kappa$-deformation from a different, complementary perspective. The starting point are the commutational relations of $\kappa$-Minkowski space (\ref{11}). They define a Lie algebra, which is called the $\sf{an}(3)$ algebra. In this name $\sf a$ stands for `abelian' (generator $X^0$) while $\sf n$ for `nilpotent' (since the generators $X^i$ are represented by nilpotent matrices), for example, the 5-dimensional representation of these matrices reads
\begin{equation}\label{5drep}
X^0 = -\frac{i}\kappa \,\left(\begin{array}{ccc}
  0 & \mathbf{0} & 1 \\
  \mathbf{0} & \mathbf{0} & \mathbf{0} \\
  1 & \mathbf{0} & 0
\end{array}\right) \quad
X^i = \frac{i}\kappa \,\left(\begin{array}{ccc}
  0 & {{\epsilon}\,{}^T} &  0\\
  {\epsilon} & \mathbf{0} & {\epsilon} \\
  0 & -{\epsilon}\,{}^T & 0
\end{array}\right),
\end{equation}
with ${\epsilon}$ being a three dimensional vector with a single unit entry, and one can check that $(X^i)^3=0$.
 It is natural to consider an associated group element
\begin{equation}\label{19}
{\sf AN}(3)\ni  \hat{e}_k \equiv   e^{i {k}_i\,
\hat {X}^i}\, e^{ik_0 X^0}
\end{equation}
This group element generalizes the plane waves (as we will see below it is a solution of deformed Klein-Gordon equation) and therefore it is called sometimes the `non-commutative plane wave' \cite{AmelinoCamelia:1999pm}. It is clear from (\ref{19}) that the momenta $k_\mu$ have a natural interpretations to be coordinates on the group manifold ${\sf AN}(3)$. Calculating the matrix form of the non-commutative plane wave with the help of the representation of the generators (\ref{5drep}) we find
\begin{equation}\label{20}
    \hat e_k = K_A{}^B= \frac1\kappa \left(\begin{array}{ccc}
  \bar P_4 & -\mathbf{P}e^{-k_0/\kappa} & P_0 \\
  -\mathbf{P} &\kappa\, \mathbf{\bbbone} & -\mathbf{P} \\
  \bar P_0 & \mathbf{P}e^{-k_0/\kappa} &  P_4
\end{array}\right)
\end{equation}
where $(P_0, \mathbf{P}, P_4)$ are given by (\ref{4}), and
\begin{eqnarray}
 \bar {P_4}(k_0, \mathbf{k}) &=& \kappa \cosh
{{k_0}/\kappa} + \frac{\mathbf{k}^2}{2\kappa}\,
e^{  {k_0}/\kappa} \nonumber\\
\bar {P_0}(k_0, \mathbf{k}) &=&\kappa  \sinh
{{k_0}/\kappa} - \frac{\mathbf{k}^2}{2 \kappa}\, e^{  {k_0}/\kappa}
\label{21}
\end{eqnarray}
The matrix $K_A{}^B$ is the matrix representation of a general ${\sf AN}(3)$ group element. To see what is the group manifold of this group let us act with this matrix on a vector in $\mathbb{R}^5$ with only fifth component non-zero
\begin{equation}\label{22}
\frac1\kappa  \left(\begin{array}{ccc}
  \bar P_4 & -\mathbf{P}e^{-k_0/\kappa} & P_0 \\
  -\mathbf{P} &\kappa\, \mathbf{\bbbone} & -\mathbf{P} \\
  \bar P_0 & \mathbf{P}e^{-k_0/\kappa} &  P_4
\end{array}\right)\, \left(\begin{array}{c}
  0 \\
  \mathbf{0}  \\
  \kappa
\end{array}\right)=\left(\begin{array}{c}
  P_0 \\
 - \mathbf{P}  \\
  P_4
\end{array}\right)
\end{equation}
We see therefore that the points on the ${\sf AN}(3)$ group manifold can parametrized by five numbers $P_0, \mathbf{P}, P_4$ satisfying
\begin{equation}\label{23}
  P_4^2-P_0^2+ \mathbf{P}^2=\kappa^2\,,\quad P_0+P_4 = e^{k_0/\kappa} >0\,,\quad P_4>0\,,
\end{equation}
which is nothing but a submanifold of the four dimensional de Sitter.

{\em The momentum space associated with $\kappa$-deformation is curved.}

It is worth stoping here for a moment, to comment on the structure of momentum space. In special relativity the momentum space is a structureless flat linear space and therefore there exists the most natural Cartesian coordinate system there and, for a simple dimensional reason, there is no real alternative to the standard, linear composition law of momenta. The situation changes dramatically when the energy scale $\kappa$ becomes available, because then we can construct arbitrary nonlinear structures. This means, in particular that there are a priori no privileged coordinates on curved momentum space, and all of them should describe the same physics, unless there is a good physical reason to prefer a particular one. This means that we could describe deformed physics in terms of any coordinates, or any deformed algebra basis. In this review we will mainly use two bases the bicrossproduct one, with momentum variables $k$ and the classical one with momentum variables $P$. There is one more convenient basis, the so-called normal one, which is briefly presented in the Appendix.

The non-commutative plane wave (\ref{19}) exhibits properties that are in one to one correspondence with the coproduct and antipode, shedding light on the meaning of them. For example, consider the product of two plane waves, i.e., the composition of two ${\sf AN}(3)$ group elements
\begin{equation}\label{24}
    \hat e_{k\oplus l} \equiv \hat e_k \hat e_l = e^{iX^i(k_i + e^{-k_0/\kappa}l_i)} e^{i X^0(k_0 + l_0)}\,,
\end{equation}
so that we see that the coproduct (\ref{8}) of momenta is nothing but the prescription following from the group elements composition rule. This is not really surprising: in the $\kappa$-deformed case the momentum space is a non-abelian group and therefore the only meaningful way to combine momenta is to make use of the group product (in fact the same holds in the standard, un-deformed case; here momenta form an abelian group with addition being the group operation.)

Similarly, the antipode (\ref{18}) can be naturally understood as an expression for the inverse group element (or the action of the adjoint)
\begin{equation}\label{25}
   (\hat e_k)^\dag = e^{-i  k_0 X^0} e^{-i   k_i X^i} = e^{-i \hat  (e^{k_0/\kappa}k_i)X^i}e^{-i  k_0 X^0} = \hat e_{S(k)}\,.
\end{equation}

Having the plane waves we can define the Fourier transform, in analogy with the standard un-deformed case. Since the momentum manifold is commutative (albeit curved), it is convenient to start with the scalar function $\phi(k)$ and define the corresponding spacetime function as
\begin{equation}\label{26}
 \phi(X) = \int_{{\sf AN}(3)} d\mu(k) \hat e_k\, \phi(k)\,,
\end{equation}
where we return to interpreting $X$ as an abstract non-commutative variable.

The measure $d\mu(k)$ (\ref{26}) is assumed to be the canonical, translational invariant measure on ${\sf AN}(3)$ manifold. To construct it, we need to find the metric, which turns out to be an easy exercise.

The metric on ${\sf AN}(3)$ is the induced one, obtained from the flat metric on $\mathbb{R}^5$
\begin{equation}\label{ds5}
 ds_5^2 = -dP_0^2 + d\mathbf{P}^2 + dP_4^2\,.
\end{equation}
Using the defining relation $P_4^2=\kappa^2+P_0^2- \mathbf{P}$ (\ref{23}) and (\ref{4}) we find
\begin{equation}\label{27}
  ds^2_{{\sf AN}(3)} = - dk_0^2 + e^{2k_0/\kappa}\, d\mathbf{k}^2
\end{equation}
which is the de Sitter space metric, as expected. Therefore
\begin{equation}\label{28}
  d\mu = \sqrt{-g(k)} d^4k = e^{3k_0/\kappa}\, dk_0 d^3\mathbf{k}\,.
\end{equation}
One can check by explicit calculation that the measure (\ref{28}) is invariant under infinitesimal Lorentz transformations (\ref{2}). This is consequence of the fact that the metric (\ref{27}) is an induced one, constructed from the manifestly Lorentz invariant metric (\ref{ds5}). In fact, expressed in classical basis the measure takes the form
\begin{equation}\label{28a}
  d\mu =\frac{d^4P}{P_4/\kappa}
\end{equation}
which is manifestly Lorentz invariant. Some more discussion concerning the $\kappa$-deformed momentum space can be found in \cite{AmelinoCamelia:2011nt} and \cite{Gubitosi:2013rna}.

This completes the preliminary description of the Fourier transform (\ref{26}). We will return to it after  defining the star product, which is a convenient tool enabling one to work with the commutative position variables $x$, instead of the non-commutative $X$.

Before doing that we must dwell for a moment on the construction of deformed differential calculus \cite{Sitarz:1994rh},
\cite{Freidel:2007yu}. Indeed, we know now how to construct functions of the non-commutative positions (\ref{26}), but we still do not know how to differentiate them. We certainly want the differential calculus to be Lorentz covariant, so that the partial derivative of  a scalar function transforms as a vector with respect to Lorentz transformations, etc. Surprisingly, the resulting differential calculus must be five-dimensional, i.e., apart from the differentials $dX^\mu$ and derivatives $\partial_\mu$ the additional objects $dX^4$ and $\partial_4$ must be included, so that the differential of the non-commutative plane wave takes the form
\begin{equation}\label{29}
 d \he_k = idX^\mu\, \hat\partial_\mu \he_k +
  idX^4\,\hat\partial_4 \he_k
  \equiv idX^A\,\hat\partial_A \he_k
\end{equation}
with
\begin{equation}\label{30}
   \hat \partial_\mu \he_k = P_\mu\, \he_k, \quad \hat \partial_4 \he_k = (\kappa-P_4)\, \he_k
\end{equation}
where $P_A=(P_\mu, P_4)$ are given by (\ref{4}). Since $P_\mu$ transform in the standard linear way as components of a Lorentz vector, while $P_4$ is a Lorentz scalar, the covariance of the calculus is manifest.

Now we have all the ingredients to introduce the star product\footnote{Here we follow the approach of  \cite{KowalskiGlikman:2009zu}. The star product associated with $\kappa$-Minkowski space is discussed also in \cite{Meljanac:2007xb}, \cite{Durhuus:2011ci}, \cite{Pachol:2015qia}.}
The idea is to use instead of the non-commutative plane waves $\he_k$ the standard exponentials on commutative spacetime with coordinates $x^\mu$ of the form $e^{i P_\mu(k) x^\mu}$, with functions $P_\mu(k)$ to be defined, such that composition and adjoint properties of $e^{i P_\mu(k) x^\mu}$ are in the one to one correspondence with the composition and adjoint rules for $\he_k$, (\ref{24}) and (\ref{25}). We fix the function $P_\mu(k)$ requiring that the action of the derivative $\hat \partial$ on $\he_k$ is $i$ times the action of the ordinary derivative on the commutative plane wave $e^{i P_\mu(k) x^\mu}$. This fixes $P(k)$ to be given by (\ref{4}), so that
\begin{equation}\label{31}
  \he_k \mapsto e^{iP_\mu\, x^\mu}\,.
\end{equation}
The star product is then defined as follows
\begin{equation}\label{32}
  \he_k\,\he_l \mapsto e^{iP(k)_\mu x^\mu}\star e^{iQ(l)_\mu x^\mu} =e^{i(P\oplus Q)_\mu x^\mu}\,.
\end{equation}
From the coproduct rule (\ref{13}) we infer
\begin{align}
(P\oplus Q)_0 &=   \frac1\kappa\, P_{0}\, Q_{+}+\kappa\,\frac{Q_0}{P_{+}}\, +\frac{\mathbf{P}\mathbf{Q}}{P_{+}}\,,
 \nonumber\\
(P\oplus Q)_i&=  \frac1\kappa\, P_{i}\, Q_{+} + Q_{i}\,,
 \label{33}
 \end{align}
 where for the later convenience we abbreviate
 \begin{equation}\label{p+}
   P_+\equiv P_0+P_4
 \end{equation}
We can also define the adjoint of the plane wave using eq.\ (\ref{25})
\begin{equation}\label{34}
   (\hat e_k)^\dag  = \hat e_{S(k)}\mapsto \left(e^{iP_\mu x^\mu}\right)^\dag = e^{i(\ominus P)_\mu x^\mu}
\end{equation}
where we used a new notation ($\ominus$) to denote the antipode.
\begin{equation}\label{35}
 ( \ominus P)_0= -P_0 + \frac{\mathbf{P}^2}{P_+}=\frac{\kappa^2}{P_+} - P_4\,,\quad (\ominus P)_i = - \kappa \frac{P_i}{P_+}\,,\quad \ominus P_+ = \frac{\kappa^2}{P_+}
\end{equation}
Notice an interesting property of the antipode: $\ominus P_\mu$ are the same functions of $P_\mu$ as $P_\mu$ is expressed in terms of $\ominus P_\mu$, exactly as it is on the case of ordinary minus. Clearly, $P \oplus (\ominus P)=0$.

This concludes our presentation of the ingredients that are necessary to construct the scalar field theory with $\kappa$-deformed symmetries.

\section{$\kappa$-deformed scalar field}

In constructing $\kappa$-deformed field theories we will start with the formulation of the theory in the non-commutative $\kappa$-Minkowski spacetime, using the star product, and then, with the help of Fourier transform, we turn to the momentum space picture. Let us therefore start with discussing Fourier transform in some details.

As it turns out the convenient definition of Fourier transform is (in what follows we do not use tilde to denote the Fourier transform; this will be either explicit or clear from the context)
\begin{equation}\label{36}
    {\phi}(P)= \frac{P_4}{\kappa}\,
\int \mathrm{d}^4x \, e^{i\ominus P x}  {\phi}(x), \quad \phi(x)= \int_{{\sf AN}(3)}
\frac{\mathrm{d}^4P}{(2\pi)^4} \frac{e^{iP x}}{P_4/\kappa}
{\phi}(P)=\int_{{\sf AN}(3)}
d\mu(P) {e^{iP x}}
{\phi}(P)
\end{equation}

Let us consider the generic quadratic term in the Lagrangian
\begin{align}
\int d^4x \, \phi(x) \star \psi^\dagger(x)&=\int d^4x \int_{{\sf AN}(3)} d\mu(P) d\mu(Q)\, \phi(P) \,\psi^*(Q)\, e^{iPx} \star\left( e^{iQx}\right)^\dagger\nonumber\\ &=\int d^4x \int_{{\sf AN}(3)} d\mu(P)  d\mu(Q)\, \phi(P) \,\psi^*(Q)\, e^{i( P\ominus Q)x}
\label{37}
\end{align}
where $P\ominus Q \equiv P\oplus (\ominus Q)$.
The expression (\ref{37}) is pretty complicated, but can be simplified considerably in a number of the following steps, making use of (\ref{33}), (\ref{35}). Consider the exponent
$$
    e^{i( P\ominus Q)x}=\exp\left[ix^0\left(\kappa\, \frac{ P_0}{Q_+} -\kappa\,\frac{Q_0}{P_+}+\frac{\kappa\, \mathbf{Q}}{P_+}\left(\frac{\mathbf{Q}-\mathbf{P}}{Q_+}\right)\right)\right]\,
    \exp\left[ix^i\left(\frac{\kappa(P_i-Q_i)}{Q_+}\right)\right]
$$
It can be simplified by the linear transformation $x^i \rightarrow x^i +x^0\, (P^i-Q^i)/P_+$ (which does not change the measure $d^4x$) to become
\begin{equation}\label{38}
 e^{i(P\ominus Q)x}=\exp\left[ix^0\, \frac{\kappa}{Q_+P_+}\left({ P_0}{P_+} -{Q_0}{Q_+}\right)\right]\,\exp\left[ix^i\, \frac{\kappa}{Q_+}\left(P_i-Q_i\right)\right]
\end{equation}
Next we use the identity $2P_0 =P_+-\kappa^2 P_+^{-1} +\mathbf{P}^2/P_+$  and the analogous one for $Q$ to rewrite
$$
 P_0P_+ - Q_0Q_+=\frac12\, (P_+^2-Q_+^2) +\frac12\left( \mathbf{P}^2-\mathbf{Q}^2\right)
$$
The second term in this expression can be removed by a linear change of position variables, as before. Then we absorb the factor into $x_0$ and change momentum variables, to obtain
\begin{equation}\label{39}
 e^{i(P\ominus  Q)x}\sim\frac{2P_+Q_+}{\kappa(P_++Q_+)} \,\left(\frac{Q_+}{\kappa}\right)^3 \exp\left[ix^0\left( P_+ - Q_+\right)\right]\,\exp\left[ix^i\left(P_i-Q_i\right)\right]
\end{equation}
where the $\sim$ indicates that the equality holds only in the integral (\ref{37}). Therefore we obtain
\begin{equation}\label{40}
\int d^4x \, \phi(x) \star \psi^\dagger(x)=\int_{{\sf AN}(3)} d\mu(P)  d\mu(Q)\,\phi(P)\,\psi^*(Q)\,\left( \frac{P_+}\kappa\right)^4 \delta\left(  P_+- Q_+\right)\, \delta^3\left(\mathbf{P}-\mathbf{Q}\right)
\end{equation}
This formula can be further simplified.
 Keeping in mind the fact that thanks to the second delta we can take $\mathbf{Q}=\mathbf{P}$, and we have
\begin{align}
 &\int dP_0\, \delta\left(  Q_+- P_+\right)\, f(P_0) = \int dP_+\, \frac{P_4}{P_+}\, \delta\left(  Q_+- P_+\right)\, f(P_0(P_+))\nonumber\\&= \frac{Q_4}{Q_+}\, f(Q_0) =  \int dP_0\,\frac{P_4}{P_+}\, \delta\left(  Q_0- P_0\right)\, f(P_0)\nonumber
\end{align}
 where we used the fact that $f(P_0(Q_+))=f(Q_0)$. Thus, finally
 \begin{equation}\label{41}
\int d^4x \, \phi(x) \star \psi^\dagger(x) =\int_{{\sf AN}(3)} d\mu(P)\, \left( \frac{P_+}\kappa\right)^3 \phi(P)\,\psi^*(P)
 \end{equation}

There is one more thing that we still have to clarify, namely the relation between $\phi^*(P)$ and $\phi(P)$ in the case of a real field. It is convenient to establish this relation first for the noncommutative plane waves, and then turn to the $P$ momentum coordinates. From (\ref{26}) we have for real field
\begin{align}
& \phi^\dag(X)= \int_{{\sf AN}(3)}e^{3k_0/\kappa}\, dk_0d^3\mathbf{k}\, \hat e_{S(k)}\, \phi^*(k) = \int_{{\sf AN}(3)} dS(k)_0d^3\mathbf{S}(k)\, \hat e_{S(k)}\, \phi^*(S(k)) =\phi(X)\nonumber\\ &= \int_{{\sf AN}(3)} e^{3k_0/\kappa}\, dk_0d^3\mathbf{k}\, \hat e_{k}\, \phi(k)\label{45}
\end{align}
and it follows that
\begin{equation}\label{46}
 \phi^*(k)= e^{-3k_0/\kappa}\, \phi(S(k))\,\mbox{ or }\, \phi^*(P)= \left(\frac\kappa{P_+}\right)^3\, \phi(\ominus P)
\end{equation}
so that, from (\ref{41}) we have
 \begin{equation}\label{41a}
\int d^4x \, \phi(x) \star \psi^\dagger(x) =\int_{{\sf AN}(3)} d\mu(P)\,  \phi(P)\,\psi(\ominus P)
 \end{equation}

 Since the $\star$-product is not symmetric the changing $\phi$ with $\psi$ in (\ref{41a}) will lead to the slightly different momentum space expression. Repeating the steps that has led us to this last equation we find
 \begin{equation}\label{41b}
\int d^4x \, \psi^\dagger(x)\star \phi(x) =\int_{{\sf AN}(3)} d\mu(P)\,\left(\frac\kappa{P_+}\right)^3  \phi(P)\,\psi(\ominus P)
 \end{equation}

It seems therefore that we have in our disposal two field bilinears which can be used to build the free field action: one resulting from (\ref{41a})
\begin{equation}\label{48}
S_{\ref{41a}}= \frac12  \int d^4x \,  \left(-\Box + m^2\right)\phi(x) \star \phi^\dagger(x)  =\frac12  \int_{{\sf AN}(3)} \frac{d^4(P)}{P_4/\kappa}\,  \phi(\ominus P)\, \phi(P)\left(P^2+m^2\right)
 \end{equation}
and another by (\ref{41b})
\begin{equation}\label{47}
S_{\ref{41b}}=  \frac12 \int d^4x \,  \phi^\dagger(x) \star \left(-\Box + m^2\right)\phi(x)  =\frac12 \int_{{\sf AN}(3)} \frac{d^4(P)}{P_4/\kappa}\,\left(\frac\kappa{P_+}\right)^3\,  \phi(\ominus P)\, \phi(P)\, \left(P^2+m^2\right)
 \end{equation}
In order to decide which of these two is correct, we must now check if they are invariant under deformed Poincar\'e transformations \cite{Agostini:2003vg}, \cite{Daszkiewicz:2004xy}. As we will see the coproduct will play an important role here.

Consider the translational symmetry first. In momentum space representation the spacetime translations correspond to phase transformations. For the infinitesimal translations in direction $\mu$ with parameter $\xi$ we have therefore
\begin{equation}\label{49}
 \delta_\mu^T\, \phi(P) = i\xi P_\mu\, \phi(P)
\end{equation}
and similarly
$$
 \delta_\mu^T\, \phi(\ominus P) = i\xi (\ominus P)_\mu\, \phi(\ominus P)
$$
If we choose to use the standard Leibniz rule we would find that the term $\phi(\ominus P)\, \phi(P)$ is not invariant, because $P_0 + (\ominus P)_0\neq 0$. At this point the coproduct magic comes to rescue. Indeed $\phi(\ominus P)\, \phi(P)$ is a product of two terms and instead of Leibnitz rule we should use the coproduct summation one. Then the result of the action of $\delta^T$ on $\phi(\ominus P)\, \phi(P)$ will become proportional to $(\ominus P \oplus P)_0 =0$ which renders this term invariant. Explicitly with the help of (\ref{13}),  (\ref{35}), and the identity $\ominus P^{-1}_+ = P_+/\kappa^2$ we find
$$
\delta_0^T\left(\phi(\ominus P)\, \phi(P)\right) = i\epsilon \left(\frac1\kappa\, (\ominus P)_0 P_+ + (\ominus P)_i P_i\, \ominus P_+^{-1}  + \kappa\ominus P_+^{-1}\, P_0\right)\phi(\ominus P)\, \phi(P)=0
$$
Since $\phi(\ominus P)\, \phi(P) = \phi(P)\,\phi(\ominus P)$ it is worth checking if this computation gives the same result for another ordering as well. The reader can check that this is indeed the case.
This shows that the action is invariant under time translations. The space translations invariance can be shown in a similar way.

Actually, the translational invariance can be elegantly derived in a bit more abstract way as follows. The bilinear lagrangian can be rewritten, by changing variables $\ominus Q \rightarrow Q$ as
$$
 L \sim  \phi(P)\phi(Q)\, \delta(P\oplus Q)\,.
$$
Therefore
\begin{equation}\label{49a}
  \delta^T_\mu L \sim \delta^T_\mu\left(\phi(P)\phi(Q)\right) \delta(P\oplus Q) = (P\oplus Q)_\mu\, \phi(P)\phi(Q)\, \delta(P\oplus Q)\,.
\end{equation}
But the last expression is of the form $x\,\delta(x)$ and therefore it vanishes identically.

Let us now turn to Lorentz invariance. Consider  rotations first. The rotation sector of $\kappa$-Poincar\'e is not deformed and therefore rotation generators satisfy the standard Leibniz rule. We have for rotation with the axis defined by the vector $\rho^i$ (in what follows, since the space is Euclidean with positive metric, we take liberty not to raise/lower indices everywhere)
$$
 \delta^R\, \phi(P) = i\rho^i \epsilon_{ijk}\, P_j\, \frac\partial{\partial P_k}\, \phi(P)
$$
and similarly
$$
\delta^R\, \phi(\ominus P) = i\rho^i \epsilon_{ijk}\, \ominus P_j\, \frac\partial{\partial \ominus P_k}\, \phi(\ominus P)
$$
Now since
$$
\epsilon_{ijk}\, \ominus P_j\, \frac\partial{\partial \ominus P_k} =\epsilon_{ijk}\, P_j\, \frac\partial{\partial P_k}
$$
both the actions (\ref{48}) and (\ref{47}) are invariant under rotations, because the measure and $P_+$ manifestly are.

Now we can turn to the invariance under boosts parametrized by the infinitesimal parameter $\lambda$. Similarly to what it was in the case of rotations we have
$$
 \delta^B\, \phi(P) = i\lambda^i \left( P_0\, \frac\partial{\partial P_i}+ P_i\, \frac\partial{\partial P_0}\right) \phi(P)
$$
and
$$
\delta^B\, \phi(\ominus P) = i\lambda^i \left(\ominus P_0\, \frac\partial{\partial\ominus P^i}+\ominus P_i\, \frac\partial{\partial\ominus P_0}\right)  \phi(\ominus P)
$$
It can be checked by a quite tedious, but direct calculations\footnote{It is easier to make this calculation in the bicrossproduct basis, as it was done in \cite{Daszkiewicz:2004xy}, and then change the basis back to the classical one.} which makes use of the boost generators coproduct
\begin{equation}\label{50}
    \triangle (N_i)=N_i\otimes \bbbone +e^{-{k_0}}\otimes N_i+\epsilon_{ijk}k_j\otimes M_k = N_i\otimes \bbbone +P_+^{-1}\otimes N_i+\epsilon_{ijk}P_j\, P_+^{-1}\otimes M_k
\end{equation}
that
\begin{equation}\label{51}
 \delta^B\,( \phi(P)\phi(\ominus P)) = i\lambda^i \left( P_0\, \frac\partial{\partial P^i}+ P_i\, \frac\partial{\partial P_0}\right)( \phi(P)\phi(\ominus P))
\end{equation}
Then the action is invariant if the measure and other terms are manifestly Lorentz invariant objects. This is the case for the action (\ref{48}) since $P_4$ is Lorentz-invariant, but {\em not} for the action (\ref{47}), because $P_+$ is not Lorentz-invariant. Therefore it is the action (\ref{48}) that should be taken as a free field action satisfying the required invariance properties. It should be remarked at this point that so far we showed the invariance under infinitesimal Lorentz transformations. As it turns out there is still a problem concerning the finite Lorentz transformations, since the defining property of the ${\sf AN}(3)$ group manifold $P_+>0$ is no Lorentz-covariant. This problem was discussed in \cite{Freidel:2007hk}, \cite{Freidel:2007yu}, and finally solved in \cite{Arzano:2009ci}.

We conclude the free field construction by presenting the momentum space action for the real scalar field interacting with an external source $J(P)$, to wit
\begin{equation}\label{52}
  S=\frac12  \int_{{\sf AN}(3)} \frac{d^4(P)}{P_4}\,  \phi(\ominus P)\, \phi(P)\left(P^2-m^2\right) + J(\ominus P)\phi(P) + J(P)\phi(\ominus P)
\end{equation}
It is worth stressing once again that this action differs from the usual one by a nontrivial measure and range of integration and by the fact that the $-P$ of the un-deformed theory is replaced by $\ominus P$. This action is the basic building block for the perturbative quantum field theory in the path integral formalism.\newline

Having derived the free field action let us now discuss briefly interaction terms, considering only the  $\phi^4$ theory. It follows from our experience with the free theory that we should start with the spacetime action being an integral of $\phi^4 =(\phi \star \phi^\dag) \star (\phi\star \phi^\dag)^\dag$ because we know that $\phi \star \phi^\dag $  transforms as a scalar. We propose therefore
\begin{equation}\label{53}
  S_{\phi^4} =\frac{\lambda}{4!}\, \int\, d^4x\, d\mu\, \phi(P)\phi^*(Q)\phi^*(R)\phi(S)\, \exp\left(i x^\mu \left[(P\ominus Q)\ominus(R\ominus S) \right]_\mu\right)
\end{equation}
where $d\mu$ denotes collectively the integration measure for all the momenta. This equation can be rewritten in a more symmetric form by integrating over $x$ and replacing $\phi^*(Q) \rightarrow \phi(\ominus Q)$ and then changing integration variables $\ominus Q \rightarrow Q$.
\begin{equation}\label{54}
  S_{\phi^4} = \frac{\lambda}{4!}\,\int \frac{d^4P}{P_4} \frac{d^4Q}{Q_4} \frac{d^4R}{R_4} \frac{d^4S}{S_4} \,\phi(P)\phi( Q)\phi( R)\phi(S)\,\delta\left(P\oplus Q\oplus R\oplus S\right)
\end{equation}
This is our final formula for the $\phi^4$ interaction term. The $\phi^3$ interactions can be constructed analogously. It should be noticed that since the field $\phi$ is commutative the product of fields in this expression imposes the symmetrization of the delta function. It should be noted that the translational invariance of this quartic interaction term is evident, because, similarly to (\ref{49a}) we have
\begin{align}
 &\delta^T_\mu\left(\phi(P)\phi( Q)\phi( R)\phi(S)\right) \delta\left(P\oplus Q\oplus R\oplus S\right)\nonumber\\ &= \left(P\oplus Q\oplus R\oplus S\right)_\mu\, \phi(P)\phi(Q)\phi( R)\phi(S)\, \delta\left(P\oplus Q\oplus R\oplus S\right)=0\,.\nonumber
\end{align}

 Further investigations concerning the $\phi^4$ $\kappa$-deformed scalar field can be found, for example in \cite{Meljanac:2010ps}, \cite{Meljanac:2011cs}.

Let us complete this section with the brief discussion of quantum scalar field theory (see \cite{AmelinoCamelia:2001fd} for an early, but still up to date discussion). What makes the $\kappa$-deformed field theory different from the standard, un-deformed one is that the integration measure over momenta is nontrivial, containing the $P_4$ factor, and that in the vertex one has the $\kappa$-modified delta $\delta\left(P\oplus Q\oplus R\oplus S\right)$ instead of the standard $\delta\left(P+ Q+ R+ S\right)$ one. It is a convenient property of the classical basis that the propagator $1/(P^2 +m^2)$ remains un-deformed. It follows that the $\kappa$-deformed, momentum space Feynman rules are analogous to the standard ones, with the new integration and vertex conservation rule used.
This observation can be probably extended to the case of higher spins, because there are good reasons to believe (see the forthcoming paper \cite{JG}) that the spin factors are not modified by $\kappa$-deformation. The discussion of gauge theories on $\kappa$-Minkowski spacetime can be found in \cite{Dimitrijevic:2014dxa}.

One can see the consequences of $\kappa$-deformation already on the free field level. For example one can ask a question if the deformation influences the short distant (ultraviolet) behavior of quantum fields, and it turns out that it indeed does. In the recent paper \cite{Arzano:2017uuh} the properties of the $\kappa$-deformed static potential have been discussed and it turned out that the potential does not diverge in the limit of the vanishing distance from the source. This suggests that $\kappa$-deformation exhibits an important property, associated with quantum gravity  that it serves as a ultraviolet cut-off for field theories. Another way of interpreting this result is that at short distances/ ultra high energies the effective spacetime dimension decreases down to D=3. Such dimensional reduction was also expected to be a property of quantum gravity \cite{Carlip:2017eud}.

\section{Conclusions}

The aim of this short review was to describe the formal structure of the $\kappa$-deformation of Poincar\'e algebra. This was done in the Hopf-algebras language in Section 2 and in the group theoretical/geometrical one in Section 3. Then these so-obtained technical tools were used to construct the theory of $\kappa$-deformed scalar field.

There are many features of $\kappa$-deformation that are still obscured. Most importantly the relation of this deformation and quantum gravity in the physical $3+1$ dimensions is still far from clear. Does the $\kappa$-deformation indeed follow from quantum gravity; if so is it an exact symmetry of flat quantum spacetime, as it is the case in $2+1$ dimensions, or is it the case in some approximation (presumably semiclassical) only?

Another interesting question concerns free $\kappa$-deformed quantum fields on nontrivial gravitational backgrounds. Does $\kappa$-deformation shed some new light on the problems and puzzles related to the gravitational theormodynamics and physics of black holes? These questions are currently being investigated.

It was recently established \cite{Arzano:2017uuh} that $\kappa$-deformation renders the point particle potential finite, serving as some kind of UV cutoff. It is interesting to ask if it acts in a similar way in the case of loop divergencies of QFT. Also is $\kappa$-deformation immune to the notorious UV/IR mixing problem of the canonically non-commutative quantum field theories?

Hopefully all these, and other, questions will find their answers in a near future.

\section{Appendix. Normal coordinates}

Even if physics cannot depend on a choice of basis, it is, of course, desirable to choose the basis being as simple as possible.
Here we describe a particularly simple basis of
$\kappa$-Poincar\`e algebra, which seemed to escape
an attention so far (with an exception of \cite{Kosinski:1999dw},
where some of the properties of this basis have been briefly
discussed.) This basis is characterized by the property that the
total momentum of two identical particles of identical momenta is
twice the individual momenta
\begin{equation}\label{a1}
    (p\oplus p)_\mu = 2p_\mu
\end{equation}
as in the undeformed case. In what follows this basis will be called the {\em normal basis} of  $\kappa$-Poincar\`e algebra, because (\ref{a1}) is the defining defining condition of normal Fermi coordinates.

Such basis is of interest from the point of view of Relative
Locality \cite{AmelinoCamelia:2011bm} (see
\cite{Kowalski-Glikman:2013rxa}  review.) Indeed, in
various constructions in the relative locality framework, for
example in the discussion of the soccer ball problem
\cite{AmelinoCamelia:2011uk} one finds the normal coordinates
satisfying eq.\ (\ref{a1}) particularly useful. Since
$\kappa$-Poinar\`e momentum space with $\kappa$-Poincar\`e (quantum)
algebra of symmetries is the most studied example of Relative
Locality it is worth constructing the associated normal basis and
investigating its properties.

 Clearly, if the momenta $p_\mu(g)$ are defined  such that a group element $g$ is represented as
\begin{equation}\label{a2}
  g=  \exp\left(ip_\mu X^\mu\right)
\end{equation}
the Fermi basis defining relation  (\ref{a1}) is satisfied. Indeed, since the
momentum composition is defined  as
\begin{equation}\label{a3}
    \exp\left(ip_{1,\mu} X^\mu\right)\exp\left(ip_{2,\mu}
    X^\mu\right)=\exp\left(i(p_{1}\oplus p_{2})_\mu
    X^\mu\right)
\end{equation}
we have
$e^{ip_\mu X^\mu} e^{ip_\mu X^\mu}= e^{2ip_\mu X^\mu}$ as required.

The commonly used basis of $\kappa$-Poincar\`e algebra is the so called bi-crossproduct basis \cite{kappaM1}, \cite{kappaM2} is
associated with the following decomposition of the group element
\begin{equation}\label{a4}
    g= e^{i{k}_i X^i}e^{i{k}^0X_0}
\end{equation}
Using Baker-Campbell-Haussdorf relation one easily checks that the relation between the bi-crossproduct and normal bases takes the form
\begin{equation}\label{a5}
    e^{ik_i X^i} e^{ik^0X_0}= e^{ip_\mu X^\mu}\,,\quad
   p_0=k_0\,,\quad p_i = k_i\frac{k_0/\kappa}{1-e^{-k_0/\kappa}}
\end{equation}
This relation can be used to derive the expressions for all the Hopf algebra structures in the normal basis from the bi-crossproduct ones (we will use the conventions of \cite{Freidel:2007hk}.) In what follows we will list these structures commenting on the properties of the physical models expressed with their help.\newline

{\em The algebra sector}
of $\kappa$-Poincar\'e contains nontrivial commutators between boosts and momenta and in normal coordinates it reads
\begin{align}
[N_i, p_0] &= i p_i\, \frac{1-e^{-p_0/\kappa}}{p_0/\kappa}\nonumber\\
[N_i, p_j] &=i\delta_{ij}\frac{p_0}{2}\, \left(1+e^{-p_0/\kappa}\right) + i\delta_{ij}\frac{\mathbf{p}^2}{2\kappa}\frac{1-e^{-p_0/\kappa}}{p_0/\kappa}+ i p_ip_j\left(\frac{1-e^{-p_0/\kappa}}{{p}^2_0/\kappa}-\frac1{p_0}\right)\label{a6}
\end{align}
This algebra looks rather complicated, but its quadratic Casimir\footnote{There is an ambiguity in defining the mass-shell condition in the case of deformed theories. The Casimir presented here corresponds to the most natural kinetic operator $\hat{\partial}$, see \cite{Daszkiewicz:2004xy} for details.} and the associated mass shell relation
\begin{equation}\label{a7}
\frac{4\kappa^2}{p_0^2}\,\sinh^2\left(\frac{{p}_0}{2\kappa} \right) \left(p_0^2-{p^2}\right)=m^2\,.
\end{equation}
are surprisingly simple.

We see that the massless particles mass-shell condition is undeformed and it follows that in the limit of ultra-high energies ($p_0\gg \kappa$) it is becoming undeformed for massive particles as well. Of course the deformation disappears in the limit of low energies $p_0\ll \kappa$, as usual. The simplicity of mass-shell relation (\ref{a7}) is the first of convenient properties of the normal basis.

Another interesting thing to notice is that in the case of a particle moving with transplackian energy $p_0\gg\kappa$ the algebra (\ref{a6}) simplifies greatly and becomes similar to the Carroll algebra, describing the world, in which velocity of light is zero,
$$
[N_i, p_0] = 0\,,\quad
[N_i, p_j] =i\delta_{ij}{p_0}+i\frac{ p_ip_j}{p_0}
$$
where the last term is present because at high energies $p_0=|\mathbf{p}|$.
\newline

{\em Co-products and momentum composition rules}.
Instead of deriving the coproduct for momenta by using the corresponding expression in the bicrossproduct basis and eq.\ ({\ref{a5})  and then read the momenta composition law from the resulting formula, we will proceed in the opposite order, which turns out to be much simpler.

As we know, in the case of $\kappa$-Poincar\'e the momentum space is a group manifold and the momentum composition rule (and, in turn, the coproduct) is defined from the group composition rule. This rule  can be found by using the following string of
equalities
$$
e^{ik_i X^i} e^{ik^0X_0}e^{il_i X^i} e^{il^0X_0}=e^{i(k\oplus l)_i
X^i} e^{i(k\oplus l)^0X_0}=e^{ip_\mu X^\mu}e^{iq_\mu
X^\mu}=e^{i(p\oplus q)_\mu X^\mu}
$$
and we get
\begin{align}
   (p\oplus q)_0 &= p_0+q_0\label{a8}\\
(p\oplus q)_i &=\left(p_i\,\frac{1}{f(p_0)} + q_i\,\frac{e^{-p_0/\kappa}}{
f(q_0)}\right) f(p_0+q_0)\,, \label{a9}
\end{align}
where
\begin{equation}\label{a9a}
  f(p_0)
\equiv\frac{p_0}\kappa\, \frac1{1-e^{-p_0/\kappa}}\,.
\end{equation}
One checks that $(p\oplus p)_\mu =2p_\mu$; indeed
$$
(p\oplus p)_i =p_i\, f^{-1}(p_0)f(2p_0) (1+ e^{-p_0/\kappa}) =2p_i
$$
as expected.

We see that the normal basis shares with the bi-crossproduct one the property that the energy composition law is not deformed, which simplifies greatly both the calculations and their physical interpretation.

Knowing the momenta composition law we can deduce now the form of the coproducts for momenta
\begin{align}
\Delta p_0 &= p_0\otimes \bbbone + \bbbone \otimes p_0\,, \label{a10}\\
\Delta p_i &= \left(p_i\,\frac{1}{f(p_0)}\otimes \bbbone + e^{-p_0/\kappa}\otimes p_i\,\frac{1}{f(p_0)}\right) f\left(p_0\otimes \bbbone + \bbbone \otimes p_0\right)\,,\label{a11}
\end{align}
where the last term in (\ref{a11}) should be understood as a shorthand notation for the corresponding Taylor expansion of the function $f$.

For completeness we present the coproducts of rotation and boost generators
\begin{align}
\Delta M_i &= M_i\otimes \bbbone + \bbbone \otimes M_i\,, \label{a12}\\
\Delta N_i &= N_i\otimes \bbbone + e^{-P_0/\kappa}\otimes N_i +\frac1{\kappa f(P_0)}\, \epsilon_{ijk}\, P_j  \otimes M_k\,,\label{a13}
\end{align}

{\em The antipode}.
On curved momentum space one can define the operation  $\ominus$, which  can be understood as a deformed `minus'. It is defined by $p(g^{-1})=\ominus p(g)$, or, equivalently by the composition law requirement $p\oplus (\ominus p) =0$ for all momenta $p$. Alternatively one uses the concept of antipode $S$ which generalizes the minus for the algebra (\ref{a6}) generators.

It is a nice consequence of construction of the normal basis that the antipode for momenta is not deformed
\begin{equation}\label{a14}
  S(p_0) = - p_0\,,\quad S(p_i) = - p_i\,.
\end{equation}
For rotation and boost generators we have
\begin{equation}\label{a15}
  S(M)_i = - M_i\,,\quad S(N)_i = - e^{p_0/\kappa}\left(N_i - \frac1{\kappa f(p_0)}\,\epsilon_{ijk}\,  p_j M_k\right)
\end{equation}

The simplicity of the antipode of translational generators makes this basis well suited to discussion of the action of discrete symmetries $P$ and $T$ (see \cite{Arzano:2016egk}).

\section*{Acknowledgment} I thank Hideki Kyono and especially Kentaroh Yoshida for their hospitality in Kyoto and for encouraging me to write this review. Thanks are due to Tomasz Trześniewski for his comments on the early version of the manuscript. Comments and suggestions from Andrzej Borowiec and Jerzy Lukierski are also greatly appreciated. This work was supported  by funds provided by the National Science Center, projects number 2011/02/A/ST2/00294 and
2014/13/B/ST2/04043.

\end{document}